\documentstyle[11pt,aaspp4]{article}

\slugcomment{To appear in the Astronomical Journal, July 2000 issue}

\lefthead{Kirkpatrick et al.}
\righthead{2MASS L Dwarfs}

\begin{document}

\title{Sixty-seven Additional L Dwarfs Discovered by the Two Micron All Sky Survey (2MASS)
\footnote{Portions
of the data presented herein were obtained at the W.M. Keck Observatory which
is operated as a scientific partnership among the California Institute of
Technology, the University of California, and the National Aeronautics and
Space Administration.  The Observatory was made possible by the generous
financial support of the W.M. Keck Foundation.}} 

\author{J. Davy Kirkpatrick\altaffilmark{2}, I. Neill Reid\altaffilmark{3},
James Liebert\altaffilmark{4}, John E. Gizis\altaffilmark{2}, Adam J.
Burgasser\altaffilmark{5}, David G. Monet\altaffilmark{6}, Conard C. 
Dahn\altaffilmark{6}, Brant Nelson\altaffilmark{2}, and Rik J. 
Williams\altaffilmark{7}}
\altaffiltext{2}{Infrared Processing and Analysis
    Center, MS 100-22, California 
    Institute of Technology, Pasadena, CA 91125; davy@ipac.caltech.edu, 
    gizis@ipac.caltech.edu, nelson@ipac.caltech.edu}
\altaffiltext{3}{Department of Physics and Astronomy, University of 
    Pennsylvania, Philadelphia, PA 19104-6396; inr@herschel.physics.upenn.edu} 
\altaffiltext{4}{Steward Observatory, University of Arizona, Tucson, AZ 85721;
    liebert@as.arizona.edu}
\altaffiltext{5}{Department of Physics, MS 103-33, California 
    Institute of Technology, Pasadena, CA 91125; diver@its.caltech.edu}
\altaffiltext{6}{U.S.\ Naval Observatory, P.O. Box 1149, Flagstaff, AZ 86002;
    dgm@nofs.navy.mil, dahn@nofs.navy.mil}
\altaffiltext{7}{Department of Astronomy, MSC 152, California Institute of
    Technology, Pasadena, CA 91126-0152}

 

\begin{abstract}

We present $JHK_s$ photometry, far red spectra, and spectral
classifications for an additional 67 L dwarfs discovered by
the Two Micron All Sky Survey.  One of the goals of this new search
was to locate more examples of the latest L dwarfs.  Of the 67
new discoveries, 17 have types of L6 or later.  Analysis of these new
discoveries shows that H$\alpha$ emission has yet to be convincingly detected
in any L dwarf later than type L4.5, indicating a decline or
absence of chromospheric activity in the latest L dwarfs.
Further analysis shows that 16 (and possibly 4 more) of the new L dwarfs
are lithium brown dwarfs and that the average line strength for those 
L dwarfs showing lithium increases until type
$\sim$L6.5 V then declines for later types.  This disappearance
may be the first sign of depletion of atomic lithium as it begins to form
into lithium-bearing molecules.  Another goal of the search was to locate
nearer, brighter L dwarfs of all subtypes.  Using absolute magnitudes for
17 L dwarf systems with trigonometric parallax measurements, we develop
spectrophotometric relations to estimate distances to the other L dwarfs.
Of the 67 new discoveries, 21 have photometric distances placing them 
within 25 parsecs of the Sun.  A table of all known L and T
dwarfs believed to lie within 25 parsecs -- 53 in total --- is also
presented.  Using the distance measurement of the coolest L dwarf known, 
we calculate that the gap in temperature between L8 and the warmest known
T dwarfs is less than 350K and probably much less.  If
the transition region between the two classes spans a very small temperature
interval, this would explain why no transition objects have yet been uncovered.
This evidence, combined with model fits to low-resolution spectra of
late-M and early-L dwarfs, indicates that L-class objects span the range
1300K $\lesssim T_{eff} \lesssim$ 2000K.  The near-infrared color-color
diagram shows that L dwarfs fall along a natural, redder extension of the
well known M dwarf track.  These near-infrared colors get progressively
redder for later spectral types, with the L dwarf sequence abruptly ending
near ($J-H$, $H-K_s$, $J-K_s$) $\approx$ (1.3, 0.8, 2.1).

\end{abstract}




%

\section{Introduction}

In 1993, the first spectrum of what would later be known as an L dwarf was
published (Kirkpatrick, Henry, \& Liebert 1993).  This object, GD 165B, had
been discovered earlier by Becklin \& Zuckerman (1988) as a resolved 
companion to a nearby white dwarf.  For several years
GD 165B remained in a class by itself.  Then, beginning in 1997,
an explosion of discoveries proved that L dwarfs are quite common in the
solar neighborhood (Reid et al. 1999; see also
Delfosse et al. 1997; Ruiz, Leggett, \& Allard 1997;
Rebolo et al.\ 1998; Kirkpatrick et al.\ 1999; Goldman et al.\ 1999; 
Mart{\'{\i}}n et al. 1999b; Fan et al.\ 2000).

Providing a historical parallel to the L dwarfs are the T dwarfs, even cooler
objects spectroscopically defined as those showing methane at $K$-band
(Kirkpatrick et al.\ 1999; hereafter referred to as Paper I).  In 1995, the
first spectrum of a T dwarf was published (Oppenheimer et al.\ 1995).  This
object, Gl 229B, had been discovered as a companion to a nearby M dwarf by
Nakajima et al.\ (1995) and remained in a class by itself for several years.
Then beginning in 1999, an explosion of discoveries proved that observable
T dwarfs have a space density comparable to that of L dwarfs (Strauss et al.\ 
1999, Burgasser et al.\ 1999, Cuby et al.\ 1999, Burgasser et al.\ 2000a, 
Tsvetanov et al.\ 2000, Burgasser et al.\ 2000c).

Despite the implied space density and the subsequent profusion of cooler, T
dwarf discoveries, the number of known L dwarfs is still
small.  Additional examples, including a larger number of late-L dwarfs
and nearer (brighter) examples of all L subtypes, are needed for further
studies including parallax measurement, luminosity and temperature 
determination, kinematics, binarity, and detailed spectroscopic analyses
related to magnetic activity, lithium frequency, atmospheric abundances,
dust formation, etc.  To this end, we present here another 67 L dwarfs found 
during follow-up of candidates selected from Two Micron All Sky Survey (2MASS) 
data.  

\section{Target Selection and Spectroscopic Confirmation}

In Paper I we searched for objects in the 2MASS data
having $J-K_s \ge 1.30$, $K_s \le 14.50$, and no optical counterpart.
This technique proved efficient in finding L dwarfs but was most sensitive to
the earliest types since such a magnitude-limited search samples
a much larger volume of space for early-type 
objects of higher luminosity than it
does for late-type objects of lower luminosity.  As a result, Paper I
contained very few late-L dwarfs.  Also,
because of the small initial survey area (only 371 sq.\ deg.) very few
brighter, closer L dwarfs were identified.  In this paper we address
both deficiencies.

To find more examples of the latest L types, we have searched for 2MASS objects
having $J-K_s \ge 1.7$, $K_s \le 15.0$,  and no optical counterpart on the 
POSS-II plates.
To find nearer examples at all L types, we have also searched for 2MASS objects
having $J-K_s \ge 1.3$ and $K_s \le 13.0$ and having either no POSS-II 
counterpart
or a counterpart implying colors of $R-K_s > 6$.  

Spectroscopic follow-up of candidates selected with these search criteria
have confirmed a few dozen new L dwarfs in the still growing database.
These are listed in Table 1 along with another dozen L dwarfs that met the
same color criteria but were fainter at $K_s$.  A final L dwarf with bright
magnitudes but bluer colors ($J-K_s = 1.25$) is also included.
In Table 1, column 1 gives the object name and columns 2-7 give the
2MASS-measured magnitudes and colors.  

\subsection{Keck Observations}

Sixty-five 
of the L dwarfs in Table 1 were confirmed in 1998 August, 1998 December, 
1999 March, and 1999 July
using the Low Resolution 
Imaging Spectrograph (LRIS; Oke et al.\ 1995) at the
10m W.\ M.\ Keck Observatory on Mauna Kea, Hawaii.  A 400 lines/mm grating
blazed at 8500 \AA\ was used with a 1$\arcsec$ slit and 2048$\times$2048
CCD to produce 9-{\AA}-resolution spectra covering the range
6300 -- 10100 \AA.  The OG570 order-blocking filter was used
to eliminate second-order light.  The data were reduced and calibrated 
using standard IRAF routines.  A 1-second dark exposure was used to remove
the bias, and quartz-lamp flat-field exposures were used to normalize
the response of the detector. 

The individual stellar spectra were extracted
using the ``apextract'' routine in IRAF, allowing for the slight curvature of
a point-source spectrum viewed through the LRIS optics and using a template
where necessary. Wavelength calibration was achieved using neon+argon
arc lamp exposures taken after each program object. Finally,
the spectra were flux-calibrated using observations of 
standards LTT 9491, Hiltner 600, LTT 1020, and Feige 56
from Hamuy et al.\ (1994).
The data have not been corrected for telluric absorption, so the atmospheric
O$_2$ bands at 6867-7000, 7594-7685 \AA\ and H$_2$O bands at 7186-7273,
8161-8282, $\sim$8950-9300, $\sim$9300-9650 \AA\ are still present in
the spectra.

\subsection{Palomar Observations}

The other two 
L dwarfs in Table 1 were confirmed using the Double Spectrograph (Oke \&
Gunn 1982)
at the 5m Hale Telescope on Palomar Mountain, California.
A 2{\farcs}0 slit and dichroic beam splitter that splits the light near 
6800 \AA\ was used.  A 316 line/mm grating was placed in the red camera for
coverage from 6800 to 9150 \AA\ at a resolution of 10 \AA.  A 300 line/mm 
grating was used in the blue camera to cover the range 3375 to 6825 \AA, but 
neither of the L dwarfs had
flux detected in the blue.  Reductions were identical to those described for
the Keck data above.

The telescope used for the spectroscopic observation of each target is
listed in column 8 of Table 1 along with the observation date in column 9
and exposure time in column 10.
Finding charts for each of these L dwarfs are shown in Figure 1.

\section{Spectroscopic Classification}

Spectral types were assigned following the guidelines established in 
Paper I.  The CrH-a, Rb-b/TiO-b, Cs-a/VO-b, and Color-d ratios, as
defined in Paper I, were measured from each spectrum.  These are tabulated
in columns 2-5 of Table 2 where the names of the L dwarfs are given in column 1.
The values in parentheses after the measured value of each ratio are the class
or range in class that most closely corresponds to that value, as judged from
the primary standards plotted in Figures 10-12 of Paper I.  For those
spectra whose ratios suggest a type earlier that L5, we have listed in
column 6 the spectral class of the primary spectrum that best fits the
\ion{K}{1} profile.  These primary spectra are the ones given in Table 6
of Paper I.  The types implied by the three CrH-a, Rb-b/TiO-b, and Cs-a/VO-b
ratios along with the type implied by either the Color-d ratio (for types
$>$L5) or \ion{K}{1} fit (for types $\le$L5) have been medianed to produce
the final spectral type.

This procedure works well except for spectra with lower signal-to-noise.
In these spectra, the narrower indices (Rb-b/TiO-b and Cs-a/VO-b) are more
prone to uncertainties due to random noise spikes.  For such cases, a by-eye
comparison to the alkali and oxide features of the primary standards is a
more reliable indicator of type.  Specifically, there are twenty-one lower
quality spectra in Table 2 where the Rb-b/TiO-b and Cs-a/VO-b ratios have
been replaced by the best fit to the 7800-8600 \AA\ region encompassing the
\ion{Rb}{1} doublet, the \ion{Cs}{1} 8521 \AA\ line, the VO band near 7900
\AA, and the TiO band at 8432 \AA.  These best fits are listed in column
7 of Table 2.

Final spectral types for each object are listed in column 8 of Table 2.
The 67 new L dwarf spectra are displayed in order of increasing L subtype
in Figure 2.  Values of the spectral ratios as a function of final spectral
class are illustrated in Figure 3 with values for the 25 L dwarfs from
Paper I shown for comparison.

\section{Spectroscopic Analyses}

The sample in Table 2 not only represents a huge increase in the number of
L dwarfs known in general, but it also includes another 17 dwarfs with types 
of L6 or later.  
This larger sample can be used to study trends that evolve with spectral type. 

\subsection{H$\alpha$ Emission}

The absence of H$\alpha$ emission in
late L dwarfs, as hinted at in Paper I, can now be reinvestigated.
Column 9 of Table 2 gives for all 67 spectra the measure
of (or upper limit to) the equivalent width of the H$\alpha$ emission
feature.  Detailed spectra near H$\alpha$ are shown in Figure 4 for those
objects exhibiting emission or possible emission.  Figure 5a shows these
equivalent widths as a function of spectral subclass.  To provide
a larger sample, the 25 L dwarfs from Paper I have also been included.
As the figure shows,
typical H$\alpha$
strengths for those early L dwarfs with emission are generally a few \AA.
H$\alpha$
lines of similar strength would be detectable for several of the late-L
dwarfs here, but none shows the line.  In fact, the latest L dwarf with 
detected H$\alpha$ emission
is the L4.5 dwarf 2MASSW J2224438$-$015852, and here the 1-\AA\ equivalent
width line is observable
only because of the spectacular signal-to-noise in this spectrum.  

Figure 5b shows as a function of L subclass the percentage of L dwarfs
having H$\alpha$ emission.  Only those spectra with sufficient
signal-to-noise to detect a line of 2 \AA\ equivalent width are included
in the computation.  For type
L0, 60\% show H$\alpha$ emission of this strength, but the percentage
drops markedly for types L1, L2, and L3.  For types L4 and L5, H$\alpha$
emission is not detected at 2 \AA\ equivalent width or greater in any of our
sample.  For many of the latest L dwarfs (L6-L8) the signal-to-noise is too
poor to exclude H$\alpha$ emission at 2 \AA\ equivalent width because
our ability to detect low-level H$\alpha$ emission is compromised by the
paucity of red photons.  That having been stated, even for those few L6-L8 
dwarfs having sufficient signal-to-noise, H$\alpha$ emission was also not 
detected.

It should also be noted that H$\alpha$ strengths for the entire ensemble of 
H$\alpha$-emitting L dwarfs are
smaller than in typical dMe stars.  Both this and the absence of measureable
H$\alpha$ in mid- and late-L dwarfs indicate a decline in
chromospheric activity throughout the L dwarf sequence.   Gizis et al.\ (2000)
have used the available data on L dwarfs with and without H$\alpha$ emission
to conclude that lack of activity may correlate with youth and substellar
nature.  Such a correlation would explain why H$\alpha$ emission is not seen in
dwarfs later than L4.5 as this subclass corresponds roughly to the temperature
at which stellar interior models predict a substellar fraction of 100\%.  In
other words, a mixture of stars and brown dwarfs is expected at earlier L 
types (where H$\alpha$ emitters and non-H$\alpha$ emitters lie), but later 
than this (where no H$\alpha$ emitters are found), all objects are expected 
to be substellar.  The seeming anti-correlation between L dwarfs with lithium
absorption and those with H$\alpha$ emission is more evidence in favor of this
conclusion.  

\subsection{Lithium Absorption}

Paper I also suggested that lithium disappears (or is much weaker) in
the latest L dwarfs, probably due to the formation of lithium-bearing
molecules (Burrows \& Sharp 1999, Lodders 1999).  
We can also reinvestigate this hypothesis. 
Column 10 of Table 2 gives for each spectrum
the measure of (or upper limit to) the equivalent width of the \ion{Li}{1}
doublet.  Detailed spectra near the lithium doublet are shown in Figure
6 for those objects exhibiting Li absorption or possible absorption.
Figure 7a shows these equivalent widths as a function of spectral subclass,
and L dwarfs from Paper I have been added to increase the sample size.

For early- and mid-L
dwarfs with detected lithium, the equivalent width increases
with later subclasses.  This is the same behavior 
seen with the ground-state doublets of the other abundant alkalis 
\ion{Na}{1} and \ion{K}{1}.  In these objects
the cooler temperatures mean that more and more of the alkali atoms are in
their neutral state.  This effect, along with the pressure broadening and
the increased column density
through which the emergent radiation passes (thanks to an atmosphere made
more transparent by the removal of overlying TiO and VO absorption),
creates stronger and stronger absorption lines.  The
main difference is that for lithium, its lower cosmic abundance precludes
the development of giant absorption troughs like those produced by sodium 
and potassium (Reid et al.\ 2000).

However, the trend of increasing line strength for
lithium, unlike sodium and potassium, reverses at types around L6.5-L7 V.  
Figure 7b shows the percentage 
of L dwarfs with detectable lithium as a
function of spectral subclass.  Only those L dwarfs with sufficient signal
to see a line of 4 \AA\ equivalent width are used in the calculation. 
The percentage of L dwarfs with strong lithium may drop for types L7 and L8
(although our statistics are still poor), but
as shown in Figure 7c, the strength of the line when detected is
much weaker than that seen for types L5 and L6.  
In other words, although lithium is
sometimes seen in the latest L dwarfs, 
its strength is greatly diminished.  At face value, this seems
to run contrary to the expectation that the latest L dwarfs should show
increased line strengths of ground state \ion{Li}{1} due to their cooler
temperatures.

We interpret this as evidence for Li depletion due to the formation of
lithium-bearing molecules.  At temperatures typical of late-M dwarfs, the
primary lithium-bearing gas is Li.  At cooler temperatures, however, lithium 
will begin to form molecules such as LiOH, LiCl, or LiF depending upon the 
physics of the gas mixture.  At the pressures expected in these objects, 
Lodders (1999) concludes that LiCl is the molecule responsible for robbing 
Li out of the atmosphere.  She also shows that Li and LiCl should have equal 
abundances near 1500-1550K with LiCl being dominant at cooler temperatures.  
Because dwarfs of type L6.5-L7 show clear signs of Li depletion in their 
spectra, we can conclude that such dwarfs have temperatures in the 1500K
realm.  A more robust temperature estimate would require observations
of a band of LiCl so that abundances of Li and LiCl could be directly
compared.  

\section{Distances}

In addition to increasing the sample size of L dwarfs particularly at the 
latest types, another goal of our survey
was to find nearer examples of all L types.
A few of the brighter objects in Table 2 already have measured trigonometric
parallaxes, a couple of which are within 10 pc of the Sun.  We can use these 
parallaxes as well as other parallaxes from Dahn (priv.\ comm.) and from the 
literature to estimate distances to the rest.  Shown in Figure
8 are plots of absolute $J$ magnitude and absolute $K_s$ magnitude as a
function of spectral class for late-M, L, and T dwarfs.  The objects plotted
here are listed in Table 3, where the object name and spectral type are given 
in columns 1-2, the reference for the trigonometric parallax measure is
given in column 3, and $M_J$ and $M_{Ks}$ values (derived from
2MASS $J$ and $K_s$ photometry) are given in columns 4-5.

A second-order least squares fit to the late-M and L dwarfs
gives the following relations between absolute magnitude and spectral
class:

\begin{eqnarray}
M_J = 11.780 + 0.198(subclass) + 0.023(subclass)^2\\
M_{K_s} = 10.450 + 0.127(subclass) + 0.023(subclass)^2
\end{eqnarray}

\noindent
where $subclass=-1$ for M9 V, -0.5 for M9.5 V, 0 for L0 V, 0.5 for L0.5 V,
etc.
These relations are valid between M9 V and L8 V and are plotted
as solid lines in Figures 8a and 8b.

With equations (1) and (2) in hand, we can now provide distance estimates to
the 62 L dwarfs lacking trigonometric parallaxes in Table 2.  Here we
compare the measured $J$ and $K_s$ magnitudes to the absolute magnitudes 
implied by plugging the spectral type into equations (1) and (2).  The 
average of the $J$ and $K_s$ distance estimates is listed in column 11
of Table 2. 

Several of these L dwarfs have measured
distances or distance estimates making them
eligible for inclusion in the Catalogue of Nearby Stars (Gliese \& Jahreiss
1991); i.e., they lie within 25 pc of the Sun.  Table 4 lists from Table 2, 
from Paper I, and from the literature all known L dwarfs that are
within or possibly within this 25-pc limit.  Also
listed are the known T dwarfs believed to be within 25 parsecs.  Table 4
is ordered by spectral type, with the early L dwarfs at the top and T
dwarfs at the bottom.  This list represents the nearest, brightest 
examples of each class and as such is the list of choice for further
follow-up studies.  Columns 1 and 2 give the object name and discovery paper.
Columns 3-5 list the spectral type and $J$ and $K_s$ magnitudes.  Column 6
gives the estimated distance computed from equation (1), and column 7 gives
the estimated distance computed from equation (2){\footnote{For T dwarfs,
distances are estimated using the method of Burgasser et al.\ (1999)}}.  For 
objects having
measured trigonometric parallaxes, columns 6-7 are left blank and the measured
distance is instead listed in column 8.

\section{The Spectroscopic Gap Between L and T Dwarfs}

In Paper I the issue was also raised as to whether the L dwarf sequence
extended to types later than L8.  We have calculated that our current
search for the latest L dwarfs covers approximately 10\% of the sky.
Despite this large survey area, however, we still have not uncovered any
L dwarf significantly later than the L8 V presented in Paper I, confirming
the conclusions of Paper I.  Objects
cooler than L8 V {\it have} been found in the 2MASS data (Burgasser et al.\
1999; Burgasser et al.\ 2000a) but these are T dwarfs.  

Spectroscopically, late-L dwarfs and T dwarfs are quite different at $J$, $H$,
and $K$ bands, T dwarfs having strong bands of CH$_4$ that are
absent in L dwarfs.  In the far red, on the other hand, L and T dwarfs are
more similar.  In this region, the weak FeH bands seen in an L8 V
can also be seen in the spectra of some T dwarfs (Burgasser et al.\ 2000b), 
and both L and T dwarfs
show H$_2$O, \ion{Cs}{1}, and \ion{K}{1} absorption (Liebert et al.\ 2000).

The near-infrared spectral differences suggest that objects intermediate 
between L8 V and 
the T dwarfs will be difficult to distinguish in the 2MASS data because they 
presumably 
would have, at the inception of CH$_4$ formation, colors intermediate 
between those of an L8 V ($J-K_s \approx 2.1$) and a typical T dwarf ($J-K_s 
\approx 0.0$).  As \S8 below describes more fully, the
2MASS L dwarf discoveries appear to have, within some cosmic scatter, a
roughly monotonic
relation of $J-K_s$ color with spectral type.  That is, there is no
evidence that the $J-K_s$ color turns bluer for the L8 V discoveries, yet no
L dwarfs with $J-K_s$ colors redder than 2.1 and with types later than L8
have been found despite exhaustive search efforts. 

On the other hand, 
the spectral similarities in the far red portion of L and T dwarf
spectra mean that colors in that region are not affected by the blue
reversal seen in the near-infrared.  A search of preliminary data
from the Sloan Digital Sky Survey (SDSS) has uncovered 7 L dwarfs (Fan et al.\ 
2000) and 2 T dwarfs (Strauss et al.\ 1999; Tsvetanov et al.\ 2000) based on 
$i^*-z^*$ colors, and it appears that the $i^*-z^*$  color may be monotonic
across the L/T border.  Specifically, $i^*-z^*$ increases from 1.8 at 
L0 V, to 2.3
at L5 V, to 2.6 at L8 V, to $\sim$4 for the 2 SDSS T dwarfs.  (The jump in 
$i^*-z^*$ color between the latest L dwarfs and the T dwarfs may be caused by
the increasing strength of the \ion{K}{1} ground-state resonance doublet, which
robs the spectrum of much of its $i^*$-band flux.  This may be the analogue
to the $\sim$1-magnitude jump in $V-I$ color noted by Reid et al.\ (2000)
between L4 V and L5 V, an effect thought to be caused by the broadening of the
ground-state \ion{Na}{1} doublet.)  Selection
for objects intermediate between types L and T would thus be unbiased using
SDSS colors.  Although the statistics is still based on small numbers,
no such objects have been uncovered despite a
successful search for dwarfs on either side of the apparent gap.

These results can be explained if the transition between dwarfs of type
L8 V and the known T dwarfs covers a small range in temperature, implying that
such intermediate objects are relatively rare.  Further observational
evidence supports this theory:  The best studied T dwarf, Gl 229B, is known
to have a temperature near 950K (Marley et al.\ 1996, Allard et al.\ 1996) 
and absolute
bolometric magnitude of 17.7 (Matthews et al.\ 1996, Leggett et al.\ 1999).  
Based on spectral
appearance, the coolest L dwarf known is probably the L8 dwarf 2MASSW
J1523226+301456 (Gl 584C).  As seen in Table 3, Gl 584C has $M_J = 15.0$, a
mere 0.4 mag brighter than the $M_J = 15.4$ value for Gl 229B.  
Based on the Tinney et al.\ (1993) measurements of $BC_J = 1.9$ for the M9 
dwarf LHS 2924 and
$BC_J=1.7$ for the L4 dwarf GD 165B, we
extrapolate to $BC_J \approx 1.3$ for L8 dwarfs like Gl 584C.  
(See also Reid et al.\ 1999.)  This implies
$M_{bol} \approx 16.3$ for Gl 584C.  Both Gl 229B and Gl 584C are brown dwarfs
and thus to first order have very similar radii (Kumar 1963).  Even to second
order, because both objects are thought to
have masses near 0.045$M_\odot$ and ages older
than $\sim$0.5 Gyr (Kirkpatrick et al.\ 2000), model 
calculations show that they should have radii that
are very nearly identical (Baraffe \& Chabrier priv.\ comm., Burrows et al.\
1997).  
From the Stefan-Boltzmann law, we can therefore deduce that the temperature 
difference between Gl 584C and Gl 229B is only $\sim$350K, thus giving
Gl 584C $T_{eff} \approx 1300$K.  

Because the available evidence suggests that the T dwarf SDSS 1624+0029 is 
warmer than Gl 229B (Nakajima et al.\ 2000; Liebert et al.\ 2000; Burgasser et
al.\ 2000b), this means that the gap between L8 V and the warmest T dwarfs is
less than 350K.  We believe that the temperature range spanned by the gap is
{\it considerably} less than 350K for two reasons: (1) Equilibrium 
thermochemistry of CO and CH$_4$ (Lodders 1999, Burrows \& Sharp 1999)
would suggest that the warmest T dwarfs extend up to 1200K or above.  (2)
Our confirmation of $\sim$100 2MASS discoveries in the $\sim$700K L dwarf
temperature range would imply a considerable number of objects falling in a
transition region as broad as 350K, since L dwarfs and 2MASS-observable T 
dwarfs are known to have similar space densities.  Even though the 2MASS
color cuts employed here and by the Burgasser search only partly cover the 
transition region, the number of objects implied by a 350K gap would suggest
that at least a few of these should have already been detected.  All have
so far escaped discovery.  We conclude, therefore, that the gap must be
{\it much} smaller than 350K and possibly even less than 100K.

One final argument in favor of a transition region spanning a very small
temperature interval can be made using the recent discovery of the bright T
dwarf 2MASSW J0559$-$1404 (Burgasser et al.\ 2000c).  This object has far
weaker methane bands than any other T dwarf, a property that Burgasser et
al.\ ascribe to warmer temperature.  This would make 2MASSW J0559$-$1404
warmer than SDSS 1624+0029, which as noted above is believed to be warmer
than 950K.  Despite the weaker methane bands in 2MASSW J0559$-$1404, however,
its near-infrared color of $J-K_s = 0.22{\pm}0.06$ is comparable to that of
other T dwarfs yet still distinctly bluer, by $\sim$1.8 magnitudes, than the 
average $J-K_s$ value for an L8 V.
Forthcoming parallax and bolometric luminosity measurements for 2MASSW
J0559$-$1404 will allow us to determine an accurate temperature, thus
allowing better observational constraints on the temperature range spanned
by the L/T transition region.

\section{L Dwarf Temperature Scale}

L0 dwarfs show weaker VO bands than late-M dwarfs presumably because vanadium 
has begun to condense out of the atmospheres of the L0 dwarfs.  Based on the
thermochemical equilibrium analysis of Lodders (1999), dwarfs of type L0
must then have temperatures near 2000K where perovskite formation begins to 
rob the chromosphere of its vanadium.  Thermochemical equilibrium calculations
by Burrows \& Sharp (1999) also support $T_{eff} \approx 2000$K for L0 dwarfs.
Fits of low-resolution red/near-infrared
spectra to atmospheric models including grain
and/or dust opacities give $T_{eff} \approx 2000-2200$K for the M9 dwarf
LHS 2924 (Jones \& Tsuji 1997; Tsuji, Ohnaka, \& Aoki 1996) and $T_{eff}
\approx 1900$ for the L2 dwarf Kelu-1 (Ruiz et al.\ 1997), again suggesting
that type L0 has $T_{eff} \approx 2000$K.

Given the temperature derivation of Gl 584C from the previous section, we can
conclude that the L dwarf sequence spans a range of effective temperature from
$\sim$1300K to $\sim$2000K.  This is in good agreement with the scale
suggested by Reid et al.\ (1999) and spans a wider range than the conservative
scale proposed in Paper I.

Basri et al.\ (2000) deduce a temperature range of 1600 to 2200K for the L
dwarf sequence by comparing high resolution observations of alkali line 
profiles to allard's
atmospheric models that include dust formation and condensation.
Using the dusty atmospheres of Tsuji and of Allard, Pavlenko et 
al.\ (2000) show that acceptable fits are provided to the far red spectra of
late-M, L, and T dwarfs if an additional opacity source (molecular/dust
absorption or dust scattering) is invoked.  These authors derive a temperature 
scale running from 1200$\pm$200K for type L7 V to 2200$\pm$200K for M9.5 V.  
The intercomparison of results shows that the largest disagreement between
scales occurs at the coolest temperatures.  Here, more
work is needed before model atmospheres produce consistent
answers when comparing two different wavelength regimes (like the far red and 
the near-infrared) or when comparing high-resolution line profile fits to 
fits of low-resolution spectral energy distributions.  Nevertheless, at this 
early stage in the study of L dwarfs, it is reassuring that different
approaches lead to similar conclusions.  Independent determinations of 
effective temperatures through direct measures of luminosities and radii are,
however, still badly needed to constrain and to check the models.

\section{Colors}

Armed with a much larger sample of L dwarfs than that presented in Paper I,
we can also reinvestigate the color space occupied by L dwarfs and the
trends of color with spectral type.  Figure 9 shows the $J-H$ vs.\ $H-K_s$
diagram for M dwarfs (solid circles), L dwarfs (open circles), and T dwarfs
(open stars).   Data for the early-M dwarfs comes from the compilation of
Leggett (1992).  Colors for late-M dwarfs and L dwarfs come from 2MASS data
(Gizis et al.\ 2000, Paper I, this paper).  Colors for T dwarfs are taken
from 2MASS (Burgasser et al.\ 1999, 2000a, 2000c) 
and from the literature (Matthews et al.\ 1996, Strauss
et al.\ 1999, Tsvetanov et al.\ 2000).  For L dwarfs having 2MASS photometric 
errors of 0.10 mag or larger in either $J$, $H$, or $K_s$, small open circles
are plotted; L dwarfs with more accurate magnitudes are plotted as
larger open circles.  Superimposed here are the dwarf track (solid line, 
obscured by the M dwarfs in the middle part of the diagram) and giant track
(dashed line) from Bessell \& Brett (1988).

This figure shows that L dwarfs lie on a red extension of the familiar
dwarf track.  Those L dwarfs with well measured colors (larger open circles)
fall in an area from, roughly, ($J-H$,$H-K_s$) = (0.8, 0.5) to (1.3, 0.8) and
have a cosmic scatter similar to that seen for the M dwarfs.  However, all
indications are that the track abruptly stops at the red end
and that slightly cooler objects are sent -- via a track on this diagram yet to
be observationally determined -- to an area near ($J-H$,$H-K_s$) = (0, 0)
where the T dwarfs lie.  Based on arguments in the previous section, this
transition
track likely covers a very small range in temperature and thus will contain
relatively fewer objects.

Figure 10 shows the average near-infrared colors of late-M through late-L
dwarfs.  Colors have been averaged into spectral class bins with half subclass
spacing.  These straight averages (solid circles) are plotted as a function of 
spectral class in Figure 10 and tabulated in Table 5.  The number of objects
contributing to each average is listed in the last column of Table 5.  Also
plotted in Figure 10 are the resulting weighted averages (open circles) where
objects with more accurately measured photometry are given higher weights
than those with poorly measured photometry.  These weighted averages should
be treated with caution however as their resulting errors underestimate the
inherent object-to-object scatter in the color measures.

The photometry of Figure 10 and Table 5 is measured by 2MASS and is taken from
Gizis et al.\ (2000), Paper I, and this paper.  Colors for types earlier than
M8 are not shown because as Gizis et al.\ (2000) show that their color-based 
selection of late-M dwarfs is biased at types earlier than this.  Two other
biases are, on the other hand, still present for this sample of L dwarfs:
First, early-L types will be biased
due to the $J-K_s \ge 1.30$ color criterion employed both in Paper I and
this paper.  This should tend to inflate the observed colors of the early-L
dwarfs relative to a bias-free sample.  This effect will be partly mitigated
by the inclusion of L dwarfs from Gizis et al.\ (2000)
since a more relaxed color cut 
was used there, but the number of L dwarfs in their sample is regrettably 
small.  Second, the additional $J-K_s \ge 1.70$ color constraint 
employed in this paper will tend to inflate the colors of mid- to late-L
dwarfs.

Despite these biases, several conclusions can still be made based on 
Figure 10:  

(1) Within the scatter of the points, $J-K_s$ color increases
monotonically from late-M through late-L spectral types and has a maximum
of $J-K_s \approx 2.1$ for the late-L dwarfs.  Even though the $J-K_s$ color 
appears
to level off near 2.0 at types of $\sim$L5 and later, the $J-K_s \ge 1.70$
bias discussed above may have artifically inflated the colors of the mid-L 
dwarfs relative to late-L dwarfs.  The structure seen at early-L
types may also be an artifact of the $J-K_s \ge 1.30$ bias having
inflated the colors of L0 and L1 dwarfs and the L0.5 bin having been
based on only 3 objects.

(2) Given the same arguments above, it also appears that the $J-H$ color
increases roughly monotonically with spectral type for L dwarfs.

(3) $H-K_s$ color also appears to increase roughly monotonically with
increasing spectral type, at least through mid-L.  At late-L types, though,
the $H-K_s$ color may turn slightly bluer.  The $H-K_s$ color, however,
covers a smaller range than the other two colors, and the dispersions are
also quite large.  If there is a blueward dip at the latest L types, it
may mean that the pressure-induced H$_2$ opacity has increased markedly
at $K_s$ band.  Tokunaga \& Kobayashi (1999) overplot the L2 dwarf
Kelu-1 with the L4 dwarf GD 165B and the L7 dwarf DENIS-P J0205.4$-$1159AB
and show that the late-L object has a flux deficit at $K$-band relative
to the other two L dwarfs, a deficit which they ascribe to
increased collision-induced absorption by molecular hydrogen.  This possible
blueward dip of $H-K_s$ color for the late-L dwarfs and (if verified) its 
cause need to be studied
further with improved photometry and near-infrared spectroscopic follow-up.

Figure 8 has shown that absolute $J$ and $K_s$ 
magnitudes are well correlated with 
L dwarf spectral subclass.  As shown in Figure 11, however, the correlation
of $M_J$ and $M_{Ks}$ with individual $J-K_s$ colors exhibits much larger
scatter than the correlation of $M_J$ and $M_{Ks}$ with spectral type.
In other words, L dwarf distance estimates derived from $J-K_s$ colors
have much larger uncertainties than those derived from spectral type. 
Part of this scatter is simply due to the uncertainties in 2MASS photometry,
typically $\pm$0.07 mag for $J-K_s$ though occasionally larger.  More 
importantly, perhaps, this figure demonstrates the intrinsic limitation of
estimating spectral types and photometric distances using colors of small
baseline.  An additional, possible reason for the larger scatter in Figure
11b and 11d is that our spectral type is derived
from far red spectra, and here dust may be playing a minimal role in 
shaping
the spectrum.  In the near-infrared, on the other hand, the presence of dust
can lead to a backwarming of the atmosphere, and this will alter the amount
of H$_2$O and H$_2$ in the photosphere.  Because both of these molecules play 
a critical role in
shaping the near-infrared spectrum, slight object-to-object variations
in their opacities can lead to differences in
near-infrared colors (Chabrier et al.\ 2000).

\section{Conclusions}

We present spectra for another 67 L dwarfs discovered during follow-up
of sources identified by 2MASS.  These together with L dwarfs from Paper
I, from Gizis et al.\ (2000), and from other surveys such as SDSS and DENIS,
bring the total of known L dwarfs to well over 100.  This sample can be used 
for a variety of follow-up 
investigations.  The presence of H$\alpha$ emission is seen to decline rapidly
from early- to mid-L dwarfs, and H$\alpha$ emission is not seen
in any L dwarf later than type L4.5.  If the lack of H$\alpha$ emission is 
an indicator of youth and/or substellarity as Gizis et al.\ (2000) suggest,
then the fraction of H$\alpha$-emitters to non-emitters at any given L
subtype may reflect the fraction of stars to brown dwarfs at that class.
Lithium absorption, when detected, is seen to increase in strength from 
early- to mid-L types, but then declines markedly after type L6.5 V.  This
turnover in lithium strength may herald the depletion of atomic lithium into
lithium-bearing molecules and as such would provide a vital clue to the 
temperature scale for L dwarfs because these reactions are expected roughly 
around 1500K.  The difference in temperature between the latest L dwarf 
and Gl 229B is calculated at $\sim$350K.  This means that the gap in
temperature between L8 and the warmest of the known T dwarfs must be 
significantly less that 350K as several of the known T dwarfs are suspected 
of being warmer than Gl 229B itself.  This also means that L dwarfs span the 
likely temperature range 1300K $\lesssim T_{eff} 
\lesssim$ 2000K.  The locus of L dwarfs in near-infrared color space is also 
shown, and distances estimates are made for all L and T dwarfs
lacking trigonometric parallax measurements.  Even at this early stage in
our investigations, researchers have identified 53 L and T dwarfs known 
(or suspected) to be within 25 parsecs of the Sun, clearly indicating that
this previously hidden population of cool objects is very large.

\acknowledgements

JDK, INR, and JL acknowledge funding through a NASA/JPL grant to 2MASS
Core Project science.  AJB acknowledges support from this grant.
JDK, JEG, AJB, BN, and RJW acknowledge the support of the Jet Propulsion
Laboratory, California Institute of Technology, which is operated under
contract with the National Aeronautics and Space Administration.  
The finder charts of Figure 1 make use of the Digitized Sky Survey
(DSS), which was produced at the Space Telescope Science Institute under
U.S.\ Government grant NAGW-2166.  The DSS itself is made possible by the
existence of the POSS-I, POSS-II, and UK Schmidt photographic surveys.
The Second Palomar Sky Survey (POSS-II) was funded by the Eastman Kodak
Company, the National Geographic Society, the Samuel Oschin Foundation, the 
Alfred Sloan Foundation, the National Science Foundation grants AST 84-08225,
AST 87-19465, AST 90-23115, and AST 93-18984, and the National Aeronautics
and Space Administration grants NGL 05002140 and NAGW 1710.  The UK Schmidt
survey was carried out at the UK Schmidt Telescope operated by the Royal
Greenwich Observatory Edinburgh with funding from the UK Science and 
Engineering Research Council, until 1988 June, and thereafter by the
Anglo-Australian Observatory.
This research has also made use of the SIMBAD database, operated
at CDS, Strasbourg, France.  JDK would like to thank the rest of the
2MASS team, without whose hard work and dedication this research would
not have been possible:  Ron Beck, Tom Chester, Roc Cutri, Diane Engler,
Tracey Evans, John Fowler, Linda Fullmer, Eric Howard, Robert Hurt, 
Helene Hyunh, Tom Jarrett, Gene Kopan, Bob Light, Ken Marsh, Howard McCallon, 
Jeonghee Rho, Mike Skrutskie, Rae Stiening, Raymond Tam, Schuyler Van Dyk, 
Bill Wheaton, Sherry Wheelock, John White, Cong Xu, and anyone else he
may have inadvertently omitted.  JDK would also like to thank assistance
at Keck by Joel Aycock, Tom Bida, Randy Campbell, Teresa Chelminiak, 
Gary Puniwai, Ron Quick,
Barbara Schaefer, Chuck Sorenson, David Sprayberry, Terry Stickel, Wayne
Wack, and Greg 
Wirth and at Palomar by Rick Burruss, Karl Dunscombe, and Skip Staples.
This publication makes use of data from
the Two Micron All Sky Survey, which is a joint project of the University
of Massachusetts and the Infrared Processing and Analysis Center, funded
by the National Aeronautics and Space Administration and the National
Science Foundation.

\clearpage

\figcaption[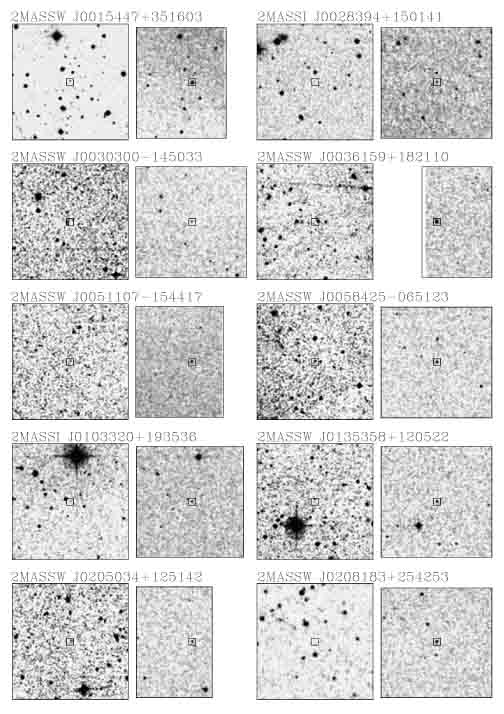,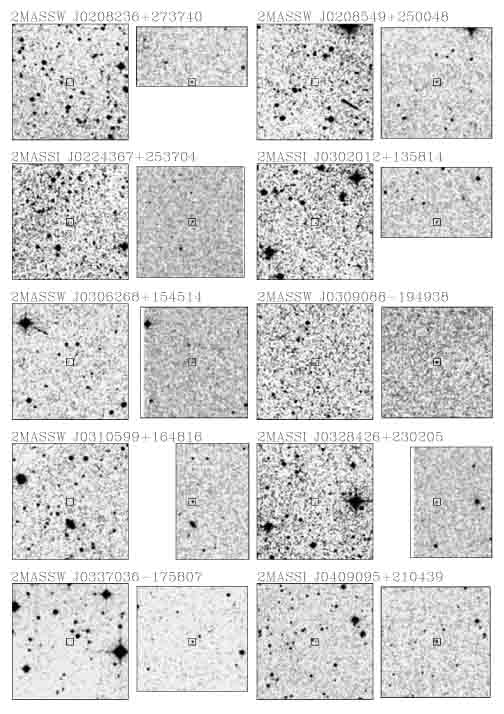,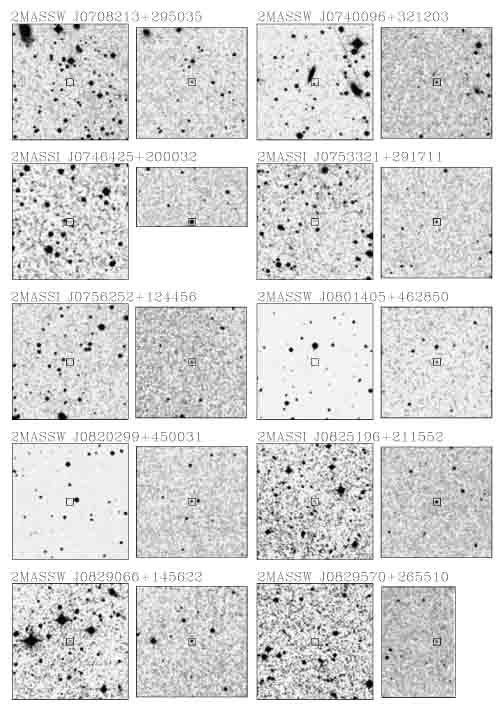,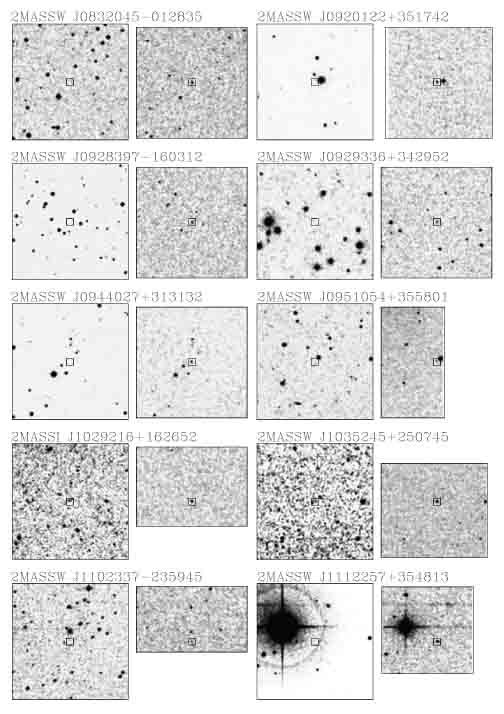,
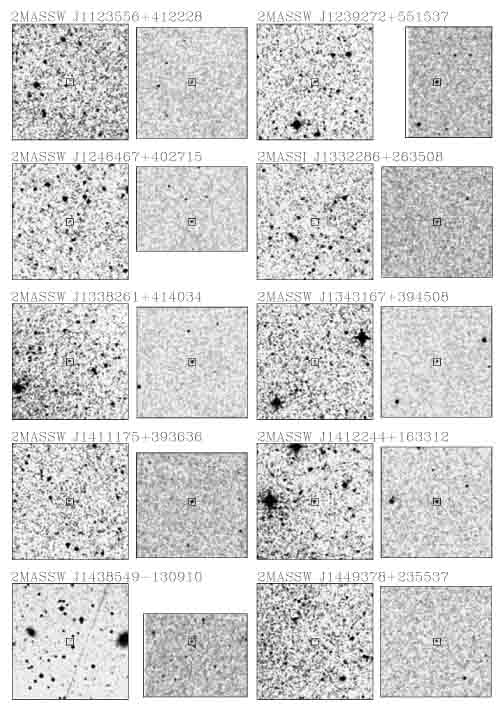,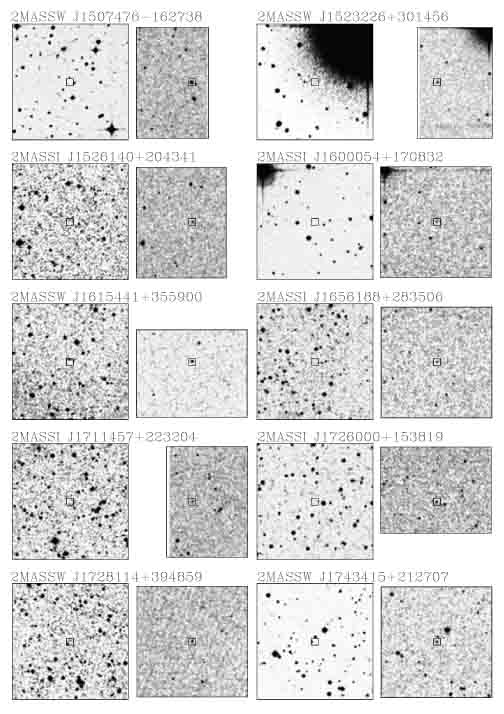,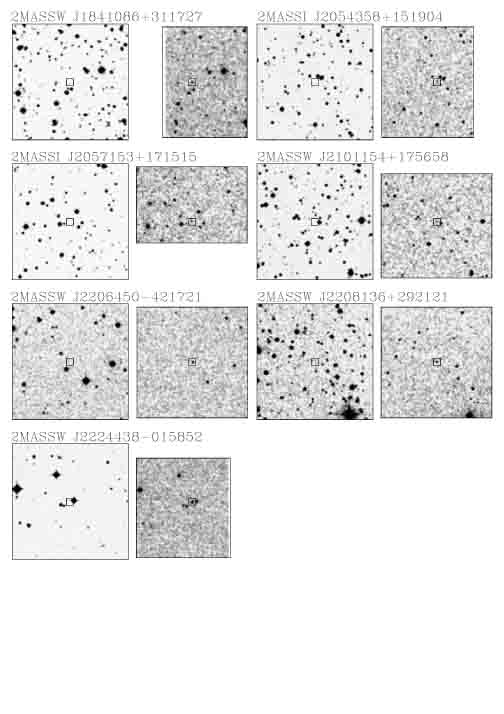]{Finder charts for each of the 
67 L dwarfs listed in 
Table 1.  For each object, two views are shown --- the Digitized Sky Survey
(DSS) image on the
left and the 2MASS $K_s$ image on the right.  Each view is to the same
scale, five arcminutes on a side with north up and east to the left.
The L dwarf is marked with a box on the 2MASS image, and a box at the same
position is also shown on the DSS image.  Note that the 
DSS image is centered on the position of the 2MASS object but that the
2MASS image does not always cover the full 5$\times$5 arcminute field.
\label{fig1}}

\figcaption[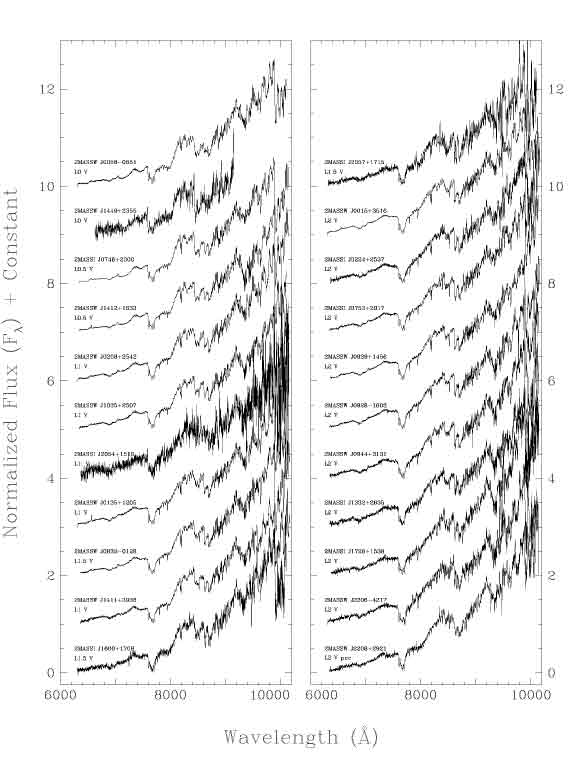,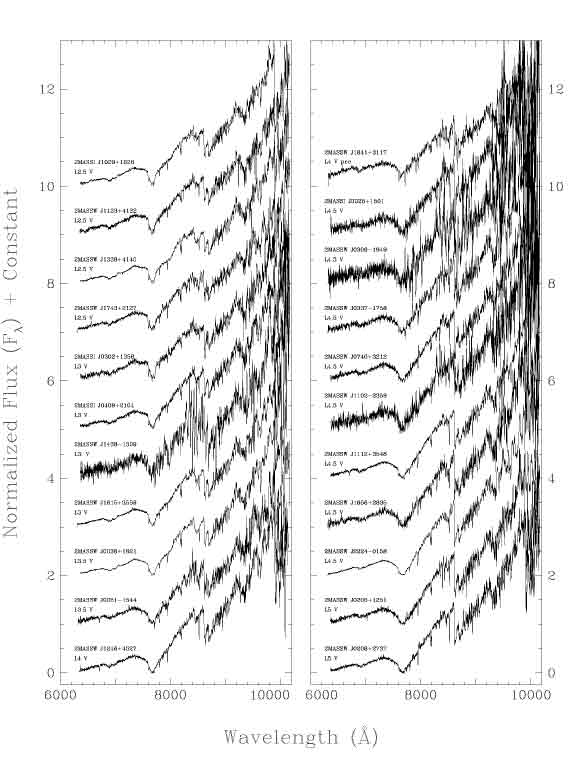,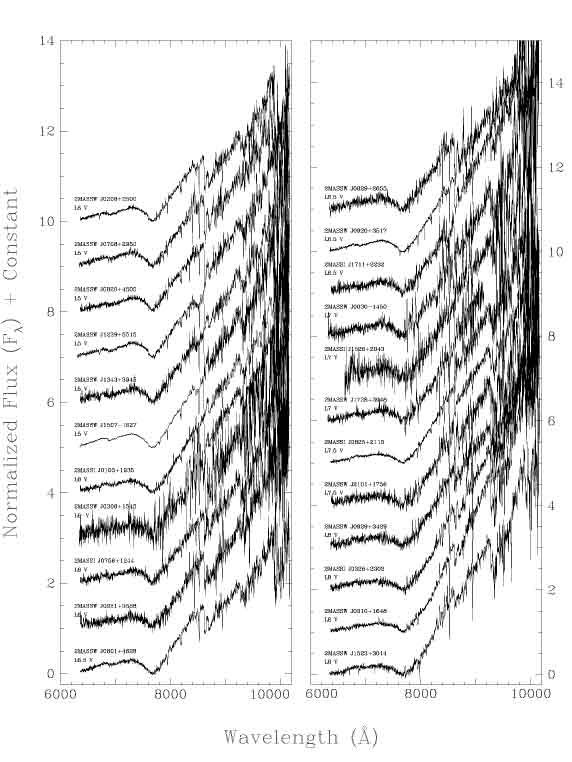]{Spectra of all 67 new 
L dwarfs.  The flux scale is
in units of $F_\lambda$ normalized to one at 8250 \AA.  Integral offsets
have been added to the flux scale to separate the spectra vertically.
Names for the 2MASS objects have been abbreviated.
\label{fig2}}

\figcaption[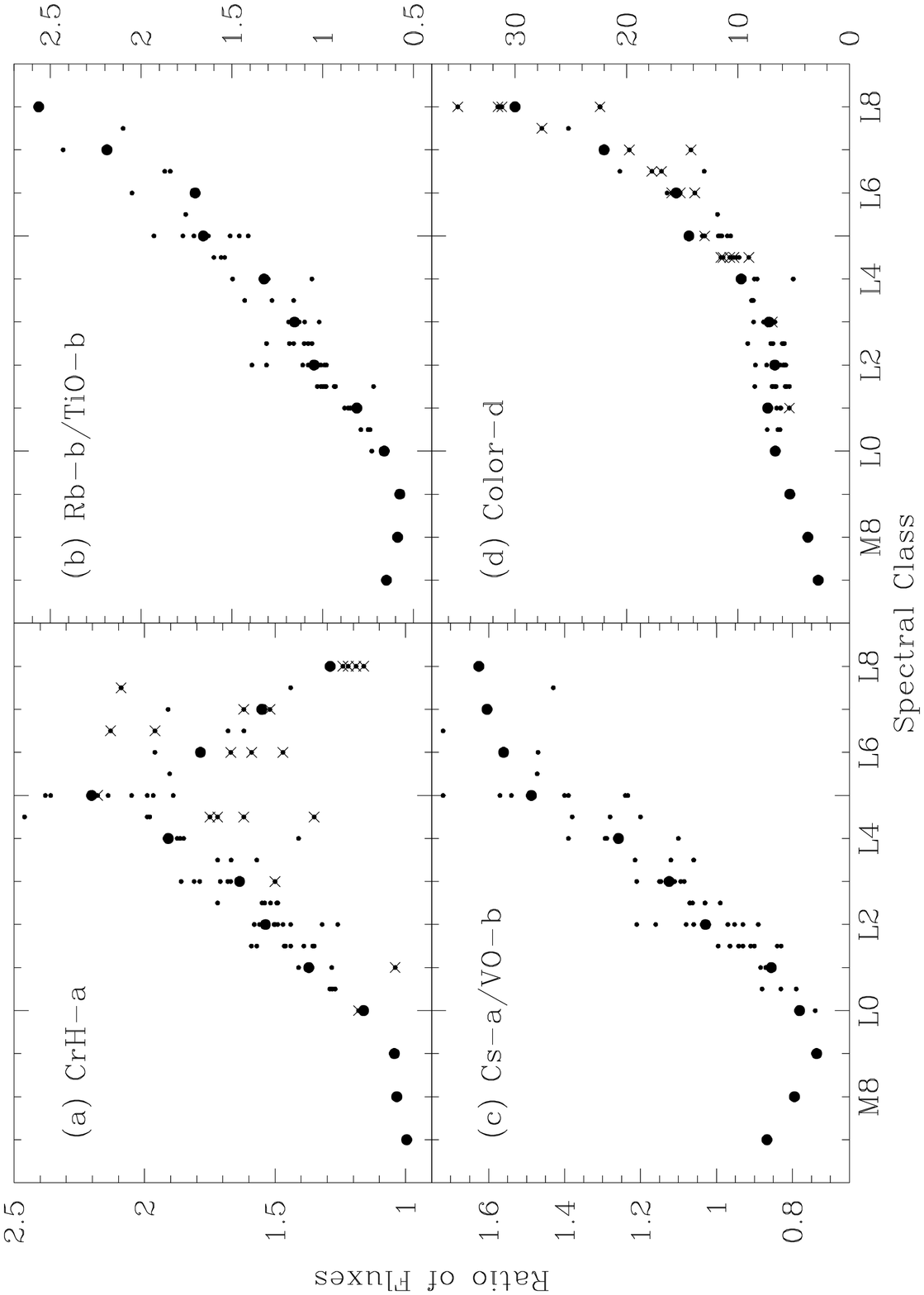]{Spectral ratios vs.\ spectral subclass for the
new L dwarfs from this paper (Table 2) together with the 25 L dwarfs 
from Paper I:
a) CrH-a, b) Rb-b/TiO-b, c) Cs-a/VO-b, and d) Color-d.  The primary 
standards from Paper I are shown as large dots; new objects from this paper
are shown with small dots.  The spectra with lower
signal-to-noise are shown with crosses in panels a) and d).
\label{fig3}}

\figcaption[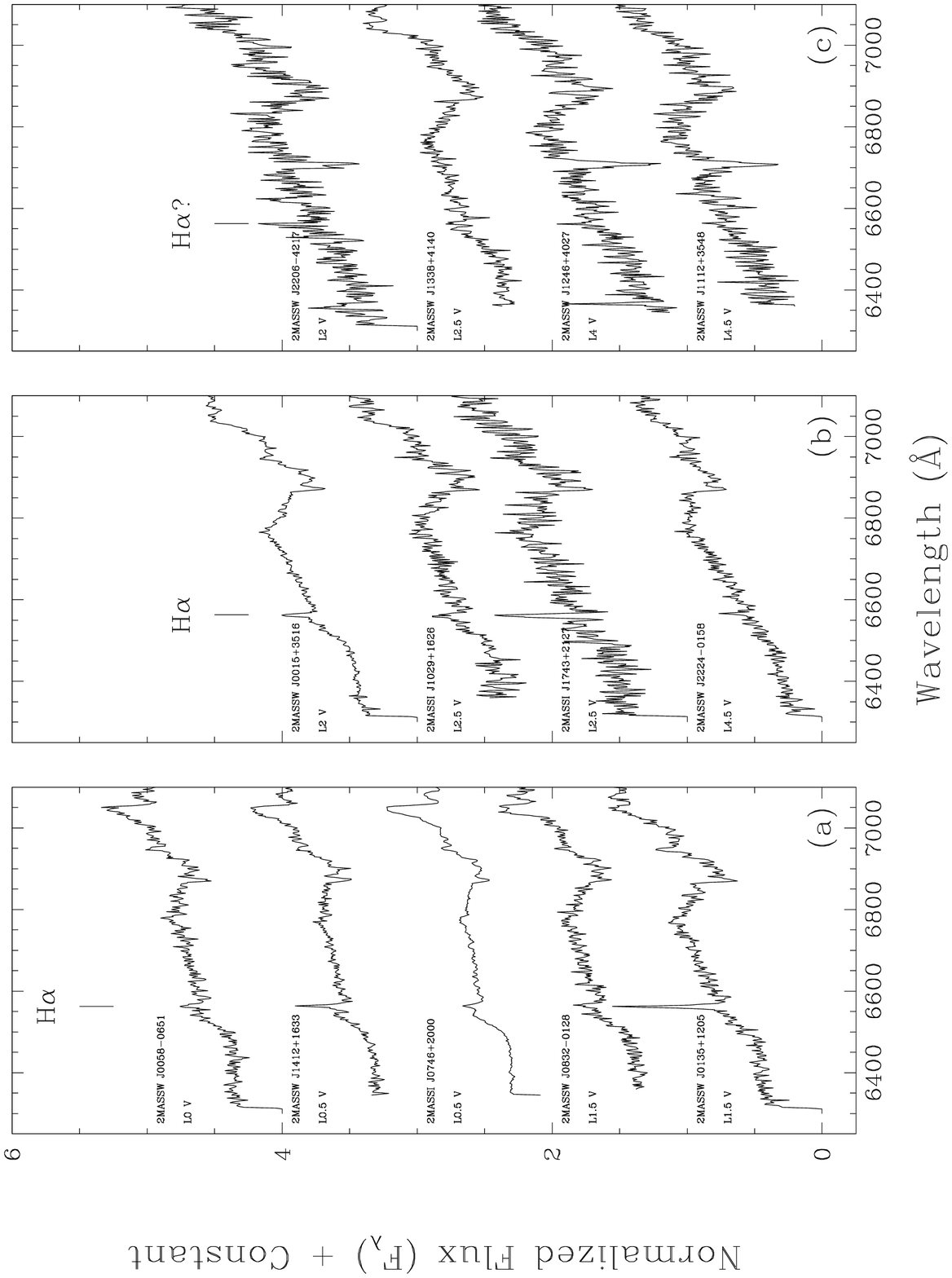]{a-b) Detailed spectra of 9 new L dwarfs showing 
H$\alpha$ emission.  c) Detailed spectra of another 4 new L dwarfs 
showing possible H$\alpha$
emission at our limit of detectability.
\label{fig4}}

\figcaption[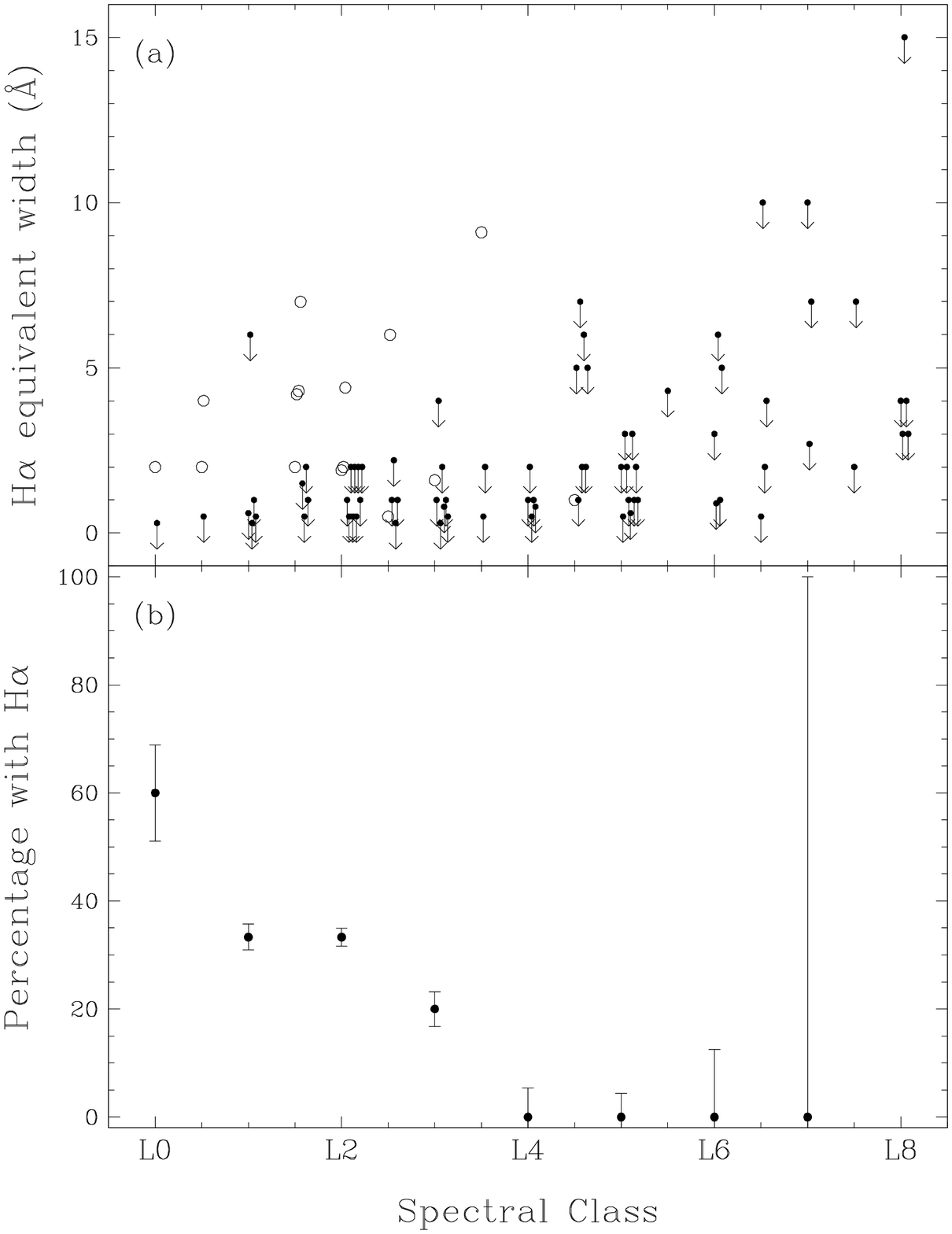]{a) H$\alpha$ equivalent widths as a function of 
spectral
subclass for the 67 L dwarfs from this paper and the 25 L dwarfs from
Paper I. Open circles denote objects having a detected H$\alpha$ emission
line.  Downward arrows denote the upper limits to the H$\alpha$ EW for those
objects where no line was detected.  To display the data more clearly,
some of the points have been given slight offsets along the x-axis.
b) Percentage of L dwarfs showing H$\alpha$ emission as a function
of spectral subclass.  The only objects used in this computation are those
where an H$\alpha$ equivalent width of 2 \AA\ or more would be detectable.
Points have been binned into integer subtypes where L0 and L0.5 dwarfs have
been combined into the L0 bin, L1 and L1.5 dwarfs combined into the L1 bin,
etc.
\label{fig5}}

\figcaption[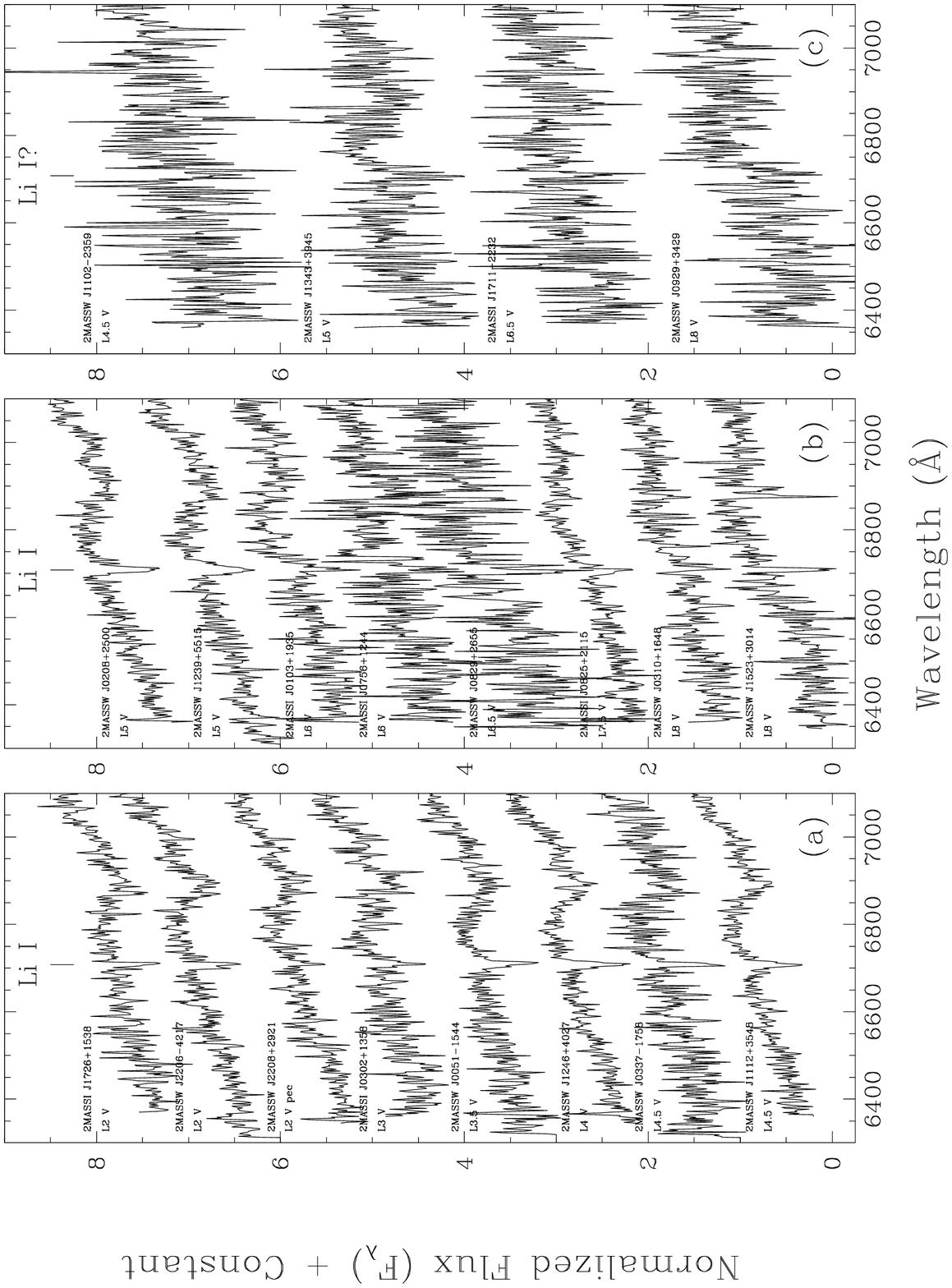]{a) Detailed spectra of 16 new L dwarfs showing 
lithium absorption.  b) Detailed spectra of another 4 new L dwarfs 
showing possible lithium
absorption at our limit of detectability.
\label{fig6}}

\figcaption[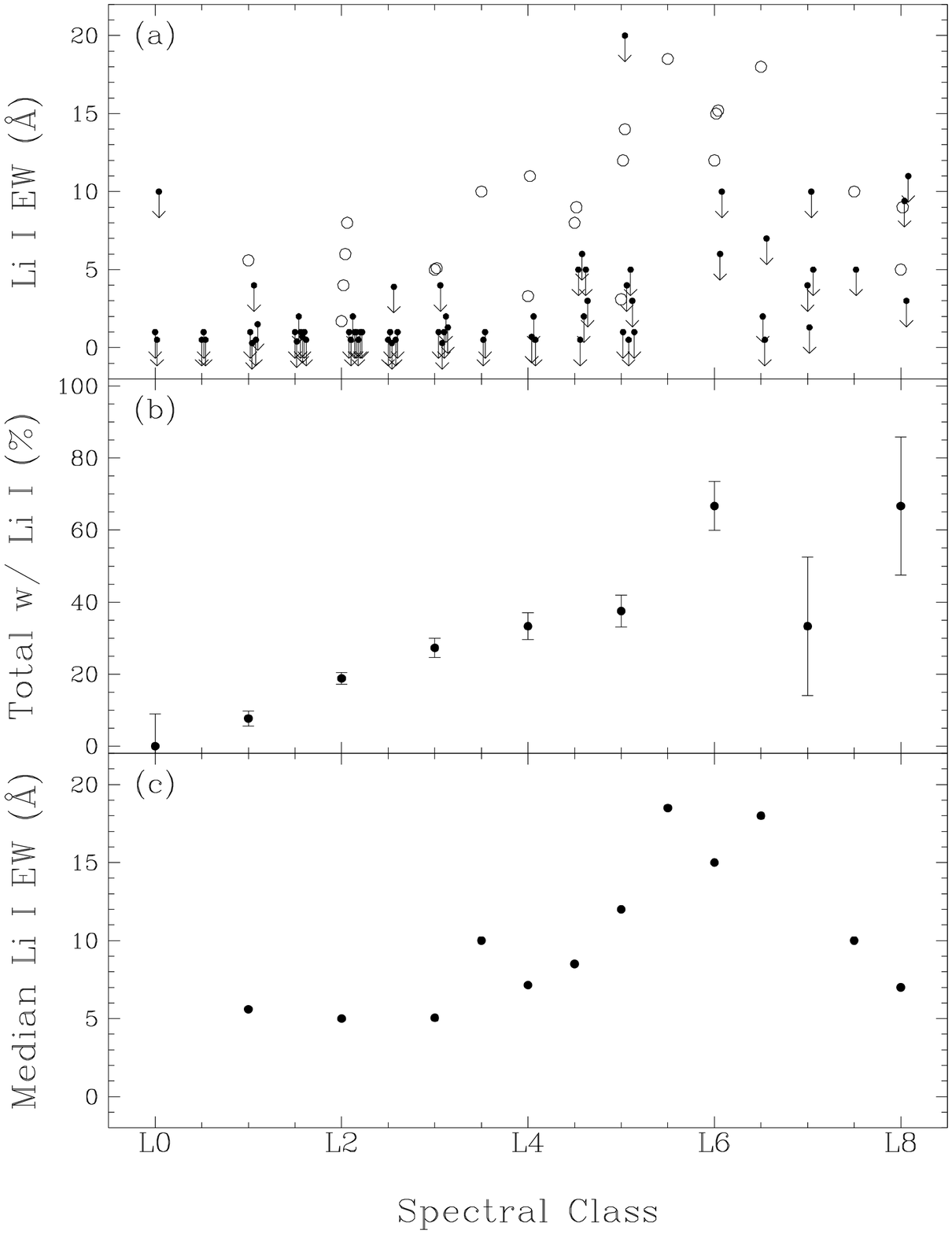]{a) Li I equivalent widths as a function of spectral
subclass for the 67 L dwarfs from this paper and the 25 L dwarfs from Paper I.
Open circles denote objects having a detected Li I absorption 
line.  Downward arrows denote the upper limits to the Li I EW for those
objects where no line was detected.  To display the data more clearly,
some of the points have been given slight offsets along the x-axis.
b) Percentage of L dwarfs showing Li I absorption as a function of 
spectral subclass.  The only objects used in this computation are those
where a Li I equivalent width of 4 \AA\ or more would be detectable.
Points have been binned into integer subtypes where L0 and L0.5 dwarfs have
been combined into the L0 bin, L1 and L1.5 dwarfs combined into the L1 bin,
etc.  c) Median Li strength as a function of spectral class for those objects
where lithium absorption was detected.  For spectral subclasses having no
objects with detected lithium, no point is plotted.  Note the drop in lithium
line strength at the latest L types.
\label{fig7}}

\figcaption[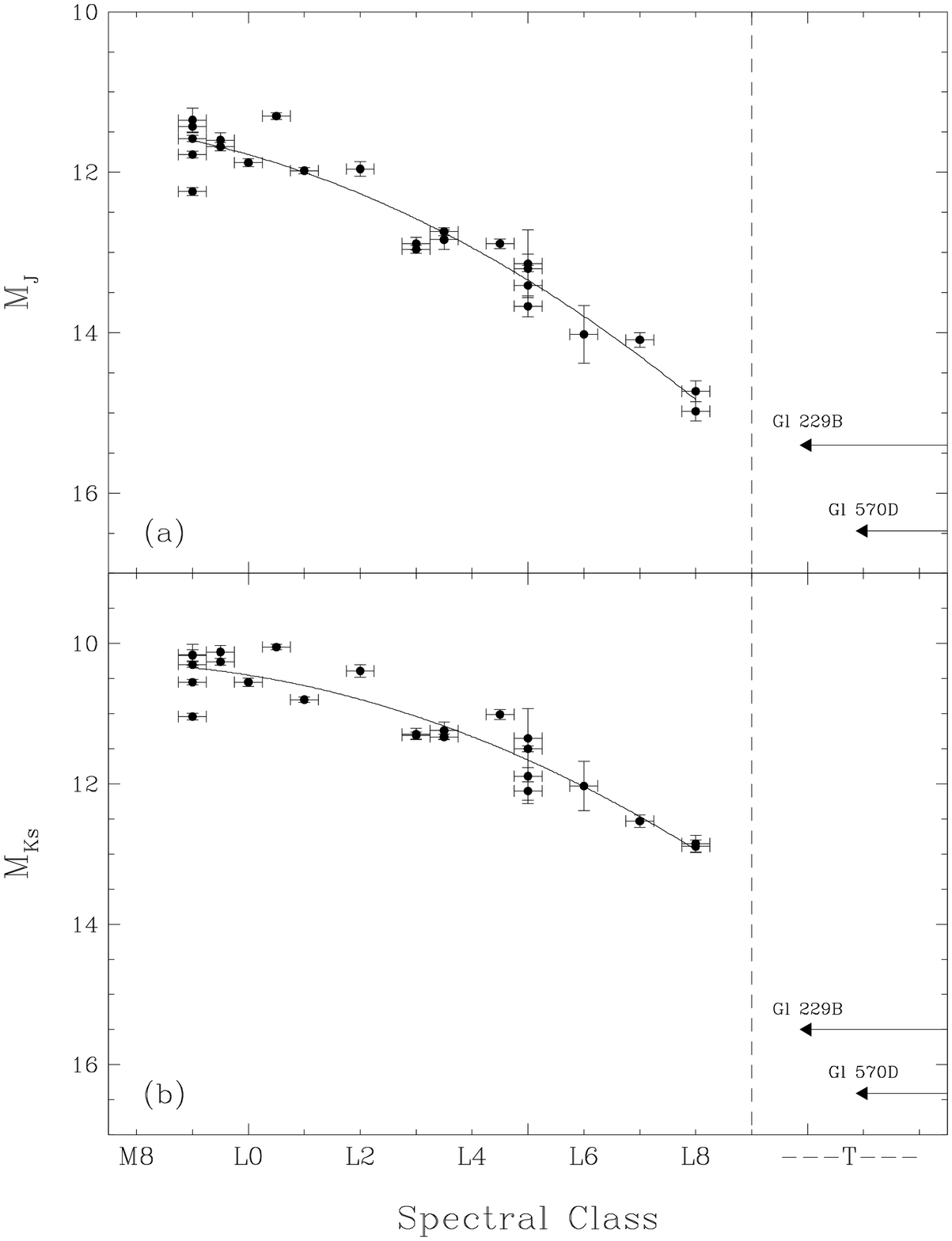]{Absolute magnitude vs.\ spectral subclass:  
a) $M_J$ vs.\
subclass, b) $M_{K_s}$ vs.\ subclass.  The second-order fits to each set of
data points (given by equations 1 and 2) are also plotted.  Absolute
magnitudes of two T dwarfs, Gl 229B and Gl 570D, are indicated by the arrows
at the bottom right of each panel.  Note that in panel a), the faintest L
dwarf (Gl 584C) is only 0.4 mag brighter at $J$ than the T dwarf Gl 229B.  
In panel b), however, this faintest L dwarf is 2.6 mag brighter at $K_s$
than Gl 229B, demonstrating the profound influence that methane has on
reshaping the near-infrared spectral energy distribution in the T dwarfs.
\label{fig8}}

\figcaption[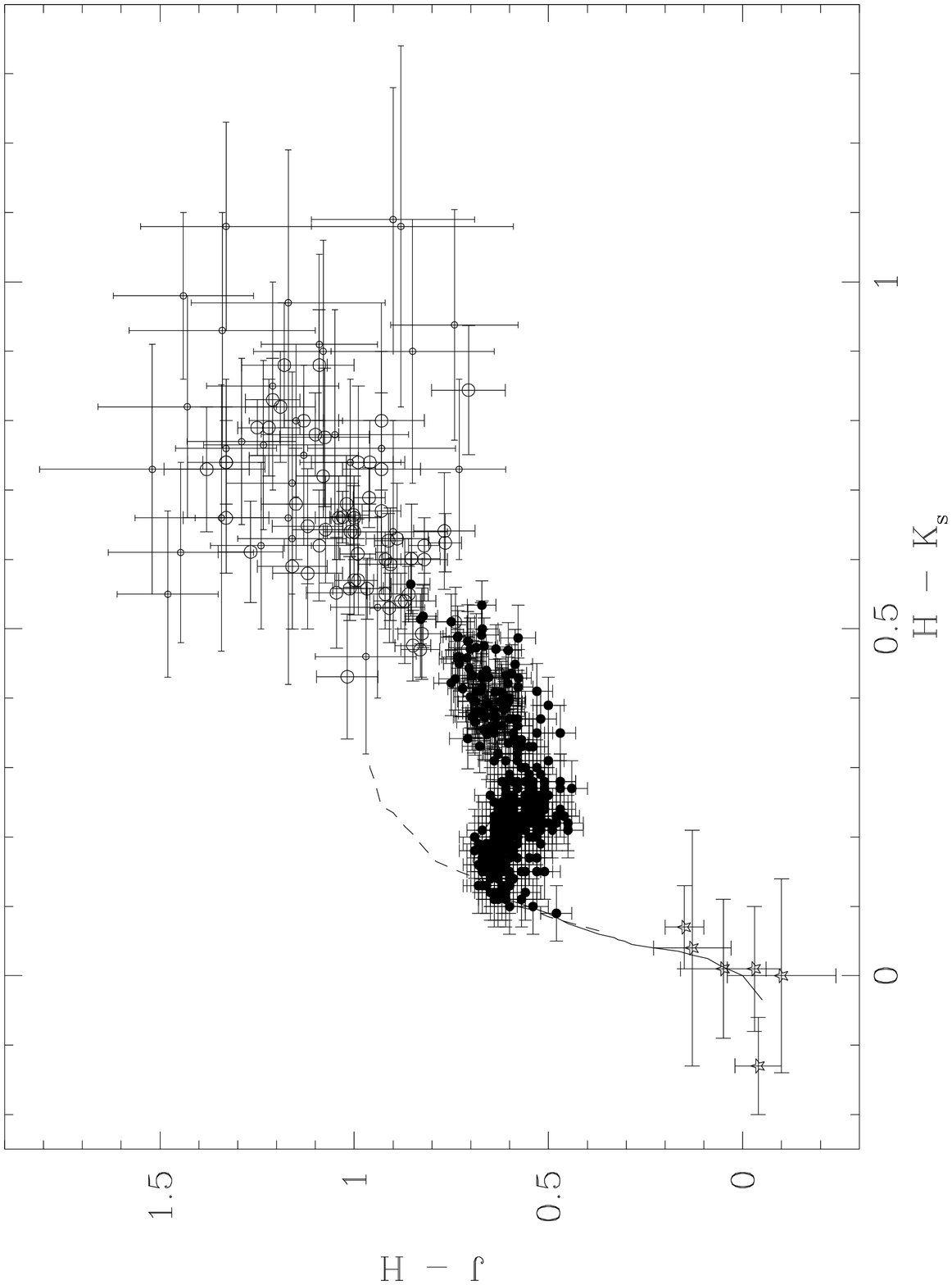]{$J- H$ vs.\ $H - K_s$ for M, L, and T dwarfs.  
M dwarfs
(solid circles) are taken from Leggett (1992) and Gizis et al.\ (2000).
L dwarfs (open circles) are taken from this paper, Paper I, and Gizis et al.
T dwarfs (open stars) are taken from Matthews et al.\ (1996), Strauss et al.\ 
(1999), Burgasser et al.\ (1999, 2000a, 2000c), and Tsvetanov et al.\ (2000).  
L dwarfs having photometric errors of 0.10 mag or larger in either $J$, $H$, or
$K_s$ are shown as small open circles, while those with better photometry are
shown as large open circles.  Tracks for dwarfs (solid line) and giants 
(dashed line) from Bessell \& Brett (1988) are also plotted.  Note that the 
latest L dwarfs lie in the upper right hand quadrant of this figure, 
diagonally opposite their slightly cooler counterparts, the 
T dwarfs at lower left.
\label{fig9}}

\figcaption[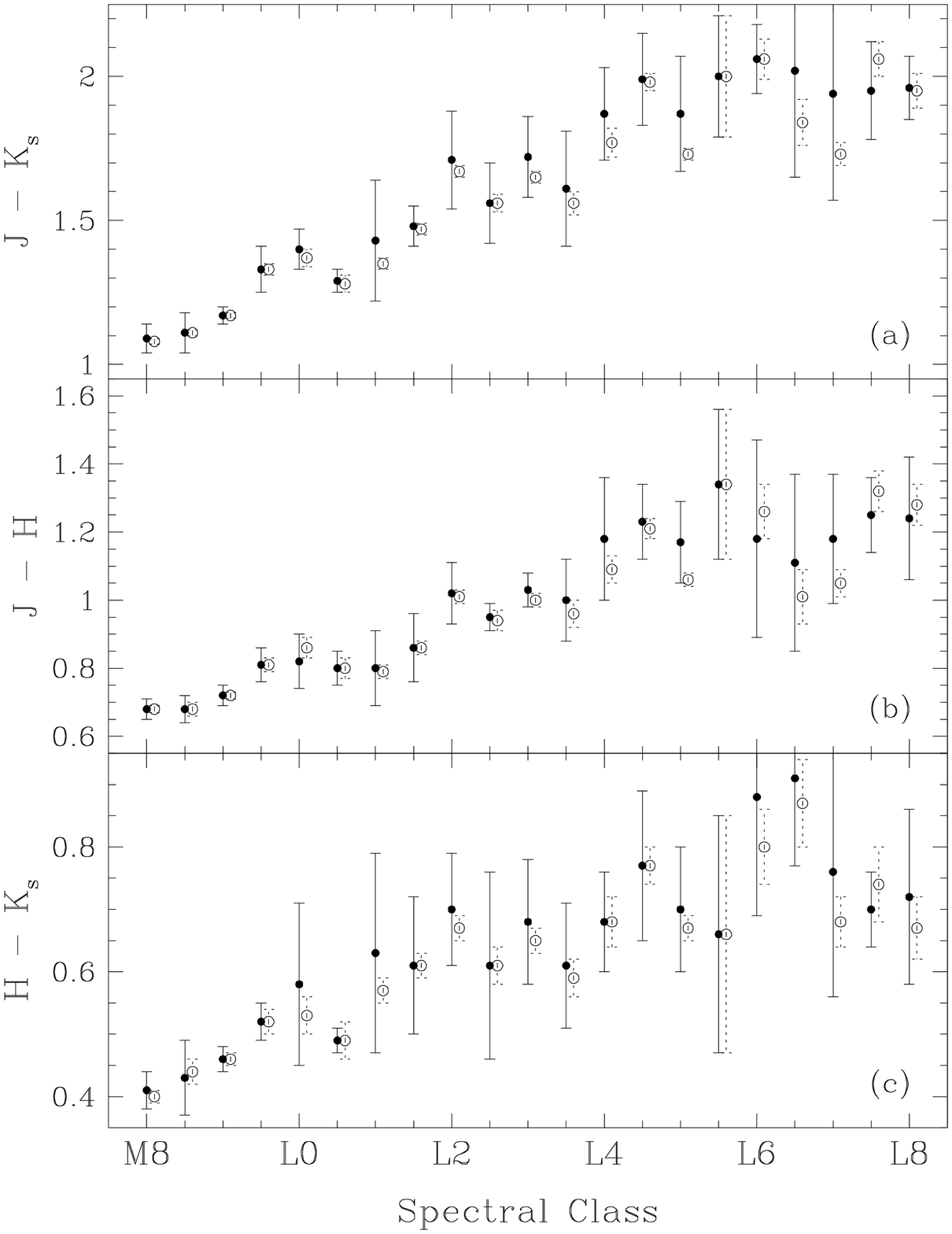]{Near-infrared color vs.\ spectral class for late-M 
through late-L dwarfs:  a) $J-K_s$, b) $J-H$, c) $H-K_s$.  Solid circles
are plotted for the averages listed in Table 5.  Open circles are weighted
averages which give a higher weight to colors with smaller measurement
errors.  See text for details.  All data shown here are from 2MASS and are
taken from this paper, Paper I, and Gizis et al.\ (2000).
\label{fig10}}

\figcaption[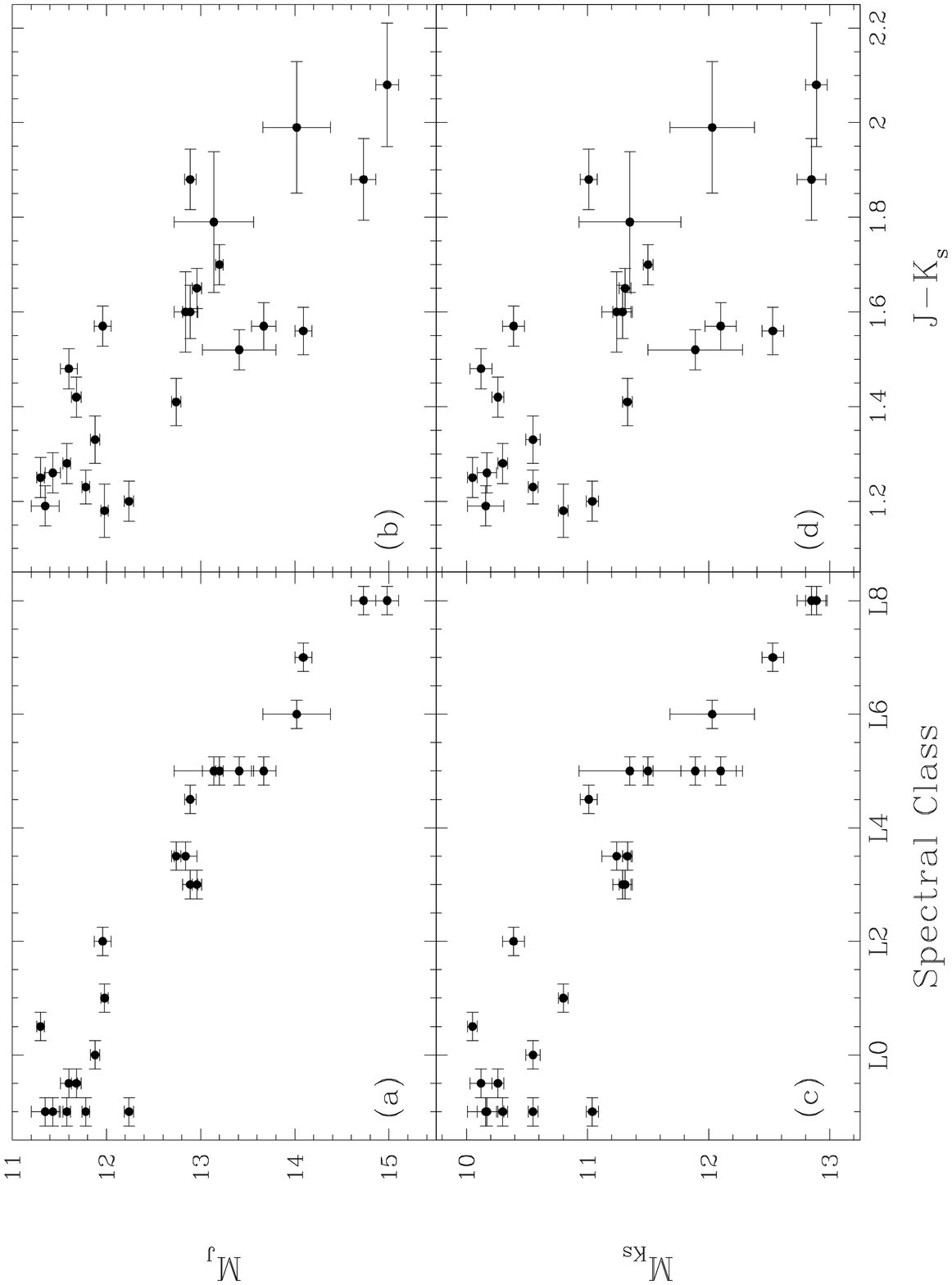]{Comparison of absolute magnitude versus spectral
type to absolute magnitude versus $J-K_s$ color for dwarfs of type M9 through
L8.  a-b) $M_J$, c-d) $M_{Ks}$, where panels a) and c) are subsections of 
Figure 8.  Note the markedly increased scatter in the color relation compared 
to the spectral type relation.  See text for details.
\label{fig11}}

%
%
%
%
%
%
%
%
%

\begin{figure}
\figurenum{3}
\plotone{kirkpatrick.fig3.ps}
\caption{}
\end{figure}

\begin{figure}
\figurenum{4}
\plotone{kirkpatrick.fig4.ps}
\caption{}
\end{figure}

\begin{figure}
\figurenum{5}
\plotone{kirkpatrick.fig5.ps}
\caption{}
\end{figure}

\begin{figure}
\figurenum{6}
\plotone{kirkpatrick.fig6.ps}
\caption{}
\end{figure}

\begin{figure}
\figurenum{7}
\plotone{kirkpatrick.fig7.ps}
\caption{}
\end{figure}

\begin{figure}
\figurenum{8}
\plotone{kirkpatrick.fig8.ps}
\caption{}
\end{figure}

\begin{figure}
\figurenum{9}
\plotone{kirkpatrick.fig9.ps}
\caption{}
\end{figure}

\begin{figure}
\figurenum{10}
\plotone{kirkpatrick.fig10.ps}
\caption{}
\end{figure}

\begin{figure}
\figurenum{11}
\plotone{kirkpatrick.fig11.ps}
\caption{}
\end{figure}

\end{document}